# Structural Phase Separation Couples to Charge-Density-Wave Formation in Kagome Metal FeGe

Boyang Zhao[1], Youngjun Ahn[1*], Qinwen Deng[2], Yidai Liu[2], Sijie Xu[4,5], Donald A. Walko[3], Stephan O. Hruszkewycz[1], Pengcheng Dai[4,5], Liang Wu[2], and Haidan Wen[1,3*]

[1]Materials Science Division, Argonne National Laboratory, Lemont, IL 60439, USA.
[2]Department of Physics and Astronomy, University of Pennsylvania, Philadelphia, PA 19104, USA.
[3]Advanced Photon Source, Argonne National Laboratory, Lemont, IL 60439, USA.
[4]Department of Physics & Astronomy, Rice University, Houston, TX 77005, USA.
[5]Rice Laboratory for Emergent Magnetic Materials and Smalley Curl Institute, Rice University, Houston, TX 77005, USA.

*youngjun.ahn@anl.gov, wen@anl.gov

**ABSTRACT**. The intertwining of charge, spin, and lattice degrees of freedom underlies the emergent properties of correlated materials. A recent prominent example is the kagome metal FeGe, which hosts coexisting charge density wave (CDW) and antiferromagnetic orders, accompanied by a lattice distortion associated with partial Ge-Ge dimerization. Using temperature-dependent high-resolution X-ray diffraction measurements, we observed a robust splitting of the lattice reflection into two coexisting peaks with distinct lattice constants at the CDW transition temperature $T_{CDW}$, providing direct evidence for a first-order structural phase transition that is absent in samples with suppressed CDW order. Furthermore, the long-range CDW order was found to be only commensurate with lattice structures with the compressed out-of-plane lattice constant. The Landau free energy analysis shows that strong lattice–charge coupling is a key factor in stabilizing long-range CDW order. Our work clarifies the critical role of structural transformation in the CDW formation and opens opportunities for strain control of electronic phases in FeGe.

The kagome lattice, a two-dimensional network of corner-sharing triangles, has emerged as a fertile playground for realizing exotic quantum states in the solid state, driven by the geometric frustration, non-trivial band topology, and electronic flat bands [1,2] that arise from kagome networks. In magnetic materials, magnetic instabilities are the hallmark of such frustrated systems, as presented by the complex spin textures in $RMn_6Sn_6$ [3] and $Fe_3Sn_2$ [4]. Further, the recent discovery of charge density wave (CDW) in kagome metal systems has unveiled a related electronic instability. In the prototypical $AV_3Sb_5$ family (A = K, Rb, Cs), CDWs emerge *via* in-plane distortions of vanadium, forming star-of-David or tri-hexagonal patterns of intertwined charge modulation and lattice distortion [5–7]. Conversely, in the bilayer system $ScV_6Sn_6$, CDW formation is driven by a soft phonon mode involving out-of-plane Sn displacements [8], leaving the vanadium kagome layer largely intact. These distinct structural motifs indicate that the formation of CDW in kagome metals inherently couples to lattice degrees of freedom.

Unlike the aforementioned materials, FeGe features the coexistence of CDW with robust antiferromagnetism [9,10], and the presence of large-amplitude Ge-Ge dimerization [11,12]. This dimerization reorganizes the Ge sublattice into alternating short and long bonds, yielding a commensurate 2×2×2 superstructure. While standard electronic mechanisms, such as Van Hove singularities near the Fermi level and Fermi-surface nesting, are relevant to the CDW formation in kagome systems [13], recent theoretical studies [14,15] and experimental observations indicate that magnetism and charge orders in FeGe are strongly intertwined. In contrast to other kagome CDW systems, which feature subtle lattice distortions, the pronounced Ge-Ge dimerization in FeGe appears to be an essential structural ingredient for realizing this intertwined state.

FeGe single crystals (space group $P6/mmm$) exhibit a CDW phase transition below $T_{CDW} \sim 100$ K [9,14,16]. The CDW emerges within the antiferromagnetic phase [9,16], which is present at temperatures up to the the Néel temperature ($T_N \sim 410$ K) [11,17]. Despite numerous reports of an abrupt and anisotropic lattice constant change across the CDW transition [18–21], the short-range CDW order (i.e., CDW correlation length at least one order of magnitudes smaller than that of the lattice) in as-grown FeGe samples complicates the underlying pictures of the role of lattice distortion in CDW formation. Moreover, post annealing studies allows control of and access to different CDW ordering lengths and intensities, depending on annealing temperatures [16,22]. Such diverse CDW ordering has led to an inconclusive picture of the transition mechanism, with studies reporting first-order characteristics [12,23] or continuous order-disorder transition [20], variant with the nature of CDW.

In this Letter, we employed temperature-dependent high-resolution X-ray diffraction to track both the fundamental lattice and CDW superlattice reflections during the CDW transition in FeGe with long-range-ordered and short-range-ordered CDWs [16]. We show that two distinct out-of-plane lattice constants coexist near $T_{CDW}$ when the CDW forms in long-range order, providing direct evidence of a first-order phase transition. In contrast, we find that FeGe with short-range and suppressed CDW order exhibits a continuous evolution of the lattice constant through $T_{CDW}$, without evidence of phase coexistence. This concurrence between CDW formation and lattice constant evolution, combined with the developed Landau theory model, suggests the strong CDW-strain coupling as an underlying mechanism to stabilize the long-range CDW order in FeGe, providing a pathway to tune intertwined orders by strain in correlated materials.

Temperature-dependent high-resolution synchrotron X-ray diffraction measurements were performed at beamline 7-ID-C of the Advanced Photon Source. A monochromatic 10 keV x-ray beam passed through a pair of vertical/horizontal slits of 50 μm and illuminated bulk single crystal samples of FeGe (~1-2 mm in size). In this work, we investigated FeGe samples annealed at two different temperature, 320 °C and 560 °C. The synthesis method and annealing conditions were reported in Ref [16]. The 320 °C -annealed FeGe exhibits a long-range CDW formation while the CDW order is highly suppressed and forms in short-range order in the 560°C -annealed one. The FeGe crystals had (0 0 1)- or (1 0 0)-type surface facets. The sample was mounted on a six-circle diffractometer equipped with a cryogenic sample stage, enabling high-resolution measurements of Bragg peak line shapes and reciprocal space intensity maps at specific reciprocal-lattice vectors as a function of sample temperature. Diffraction peaks were captured via rocking curve scans (fine scans of the X-ray incident angle) across angular peak centers using a pixelated area X-ray detector (Dectris EIGER2 X 500K). These data were converted into hexagonal $Q_x$, $Q_y$, $Q_z$ reciprocal space maps (RSMs) using the software rsMap3D [24].

The crystalline structure of FeGe consists of planar $Fe_3Ge$ layers interleaved along the *c*-axis with layers of Ge (**Fig. 1a**). Below the Neel temperature ($T_N$) and above the spin density wave temperature ($T_{SDW}$), magnetic moments localized at Fe form antiferromagnetic (AFM) order. The moments align ferromagnetically within a given $Fe_3Ge$ layer, but antiferromagnetically from one $Fe_3Ge$ layer to the next along the *c*-axis (**Fig. 1a**). Below $T_{CDW}$, CDW emerges and coexists with the AFM, accompanied by partial Ge-Ge dimerization along the *c*-axis. Approximately 1/4 of the Ge1 (denoted as Ge1d and highlighted in green in **Fig. 1c-d**) shift along the *c*-axis toward an adjacent Ge1d, forming Ge-Ge dimers.

In a 320 °C annealed FeGe crystal, both lattice (integer *h*, *k*, *l*) and CDW (half-integer indices) reflections underwent substantial changes in intensity, line shape, and peak position as the sample temperature crossed $T_{CDW}$, as shown in **Fig. 1e**. The 0.5 0 1.5 and 0.5 0 2 CDW superstructure reflections emerged below $T_{CDW}$, with peak widths comparable to the 0 0 2 lattice reflection, signifying the presence of long-range-ordered CDW. Across the same temperature range, the 0 0 2 lattice reflection displayed a splitting of the Bragg reflection into two distinct peaks at $2\theta = 35.61°$ and $35.66°$. The peak at higher $2\theta$ began to emerge at T = 96 K, coinciding with the onset of the CDW reflections and indicating a direct correlation with CDW formation. Conversely, the peak at lower $2\theta$ was completely suppressed below 92 K, where the intensity of CDW reflections saturated. Across the transition temperature range from 96 K to 92 K, the positions and widths of both peaks remained relatively constant, as indicated by the dashed lines.

To quantify the phase transition behavior, the integrated intensities were extracted as a function of temperature during both heating and cooling, as shown in **Fig. 2a**, yielding consistent transition temperatures across the 0 0 2, 1 0 2, 0.5 0 2, and 0.5 0 1.5 reflections. These results from a (0 0 1)-oriented FeGe crystal were consistently reproduced in a second sample annealed at 320 °C with a (1 0 0)-oriented surface, as evidenced by the measurements of the 0.5 0 2, 1 0 1.5, and 1 0 2 reflections (**Fig. S1**). One inference from this observation arises from the fact that the Ge1d atomic sites, which undergo substantial displacement due to dimerization, are located at the edges of the *P*-6*m*2 [12] unit cell in the *a-b* plane (**Figs. 1c-d**). This leads to a 2×2 in-plane doubling of the *P*6/*mmm* [12] parent structure (**Figs. 1a-b**). Meanwhile, Ge dimerization also doubles the unit cell along the *c*-axis, as illustrated in **Fig. 1c**. The coincident transition temperatures observed at the CDW reflections, especially between 0.5 0 2 and 1 0 1.5, therefore indicate the concurrent onset of both in-plane and out-of-plane CDW order.

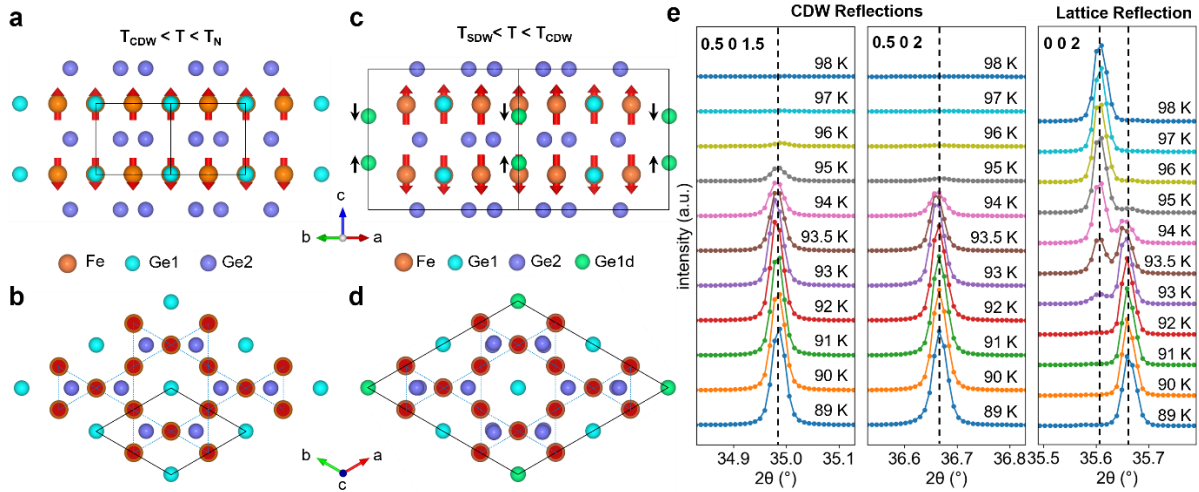

**Figure 1. Charge density wave (CDW) transition of FeGe.** The structure of FeGe in the AFM state is shown above $T_{CDW}$ in (**a-b**), and below $T_{CDW}$ in (**c-d**). The structure consists of Fe$_3$Ge layers (containing atoms labeled Fe, Ge1) and Ge$_2$ layers (all labeled Ge2); Magnetic moments are qualitatively depicted by red arrows, forming kagome lattices (dotted lines) in the *a-b* plane. The 1/4 dimerized Ge1 atoms in the Fe$_3$Ge layer are colored green and named after Ge1d. The black arrows indicate the Ge dimerization direction. The unit cells, *P*6/*mmm* vs *P*-6*m*2, are marked in solid lines. (**e**) Synchrotron X-ray diffraction intensity is integrated on the 2D detector for CDW reflections 0.5 0 1.5, 0.5 0 2, and lattice reflection 0 0 2, as a function of temperature on a (0 0 1) out-of-plane FeGe. CDW transition observed near 93.5 K, above which the CDW intensity diminishes. Concurrently, a clear double-peak profile splits 0 0 2 reflection into two peaks near $T_{CDW}$, which is absent in CDW reflections.

To understand the intensity changes observed in both the CDW and lattice reflections, we performed structure-factor calculations based on a structural model of the dimerization magnitude in the CDW phase of FeGe. The structural variations we tested were limited to applying a uniform displacement magnitude to Ge1d atoms participating in dimerization (details in Supplementary Information Section 2 and **Figs. S2e-f**). The relative changes of peak intensities calculated using Ge1d displacements of 0 Å (no dimerization) and 0.68 Å (reported in Ref [12]) provided a structure verification with the experimentally measured intensity change of the 0 0 2, 1 0 2, 0.5 0 2, and 0.5 0 1.5 above and below $T_{CDW}$.

Three key factors from our experiment were captured by the structural model with 0.68 Å Ge1d displacement: 1) the 0 0 2 peak intesity increased by 10% in the experiment from the CDW to the parent phase, compared to 15% in the model; 2) the 1 0 2 peak decreased by ~40% in the experiment, compared to a 50% decrease in the model; and 3) the near-unity intensity ratio (0.97) between the 0.5 0 2 and 0.5 0 1.5 peaks throughout the transition is only closely reproduced by 0.68 Å displacement (**Fig. S2f**). To reduce the above-noted discrepancies between experiment and structure factor calculations, an occupancy disorder [25] of the Ge1d lattice sites was proposed to reflect a mixture of Ge1d sites that do and do not dimerize. It was found that the best match to the observed intensity changes for the 0 0 2, 1 0 2, 0.5 0 2, and 0.5 0 1.5 reflections was obtained with 0.7 of Ge1d exhibiting 0.68 Å displacements, and the remaining 0.3 being zero. Such an occupancy model was also previously adopted in the refinement of the FeGe crystal structures [20,26,27].

The prominent peak splitting of the 0 0 2 reflection in **Fig. 1e** indicates the coexistence of phases with distinct lattice parameters. We therefore constructed reciprocal space maps (RSMs), shown in **Fig. 2b** and **Fig. S3**. The discrete peak splitting enabled quantification of the volume fractions of the low-temperature (LT) and high-temperature (HT) phases as a function of temperature by fitting the peaks separately at their respective positions. The resulting lattice constants as a function of temperature are shown in **Fig. 2c**, revealing an abrupt expansion from $c_{LT}$ to $c_{HT}$ of 0.15%. In contrast, the lattice constant along $Q_x$ contracts by a smaller amount of 0.04%, as determined from the RSM of the 1 0 2 reflection (**Fig. S3**). This is in good agreement with the ~0.05% contraction of the in-plane lattice constants reported in neutron Larmor diffraction studies of similarly annealed samples [17,18].

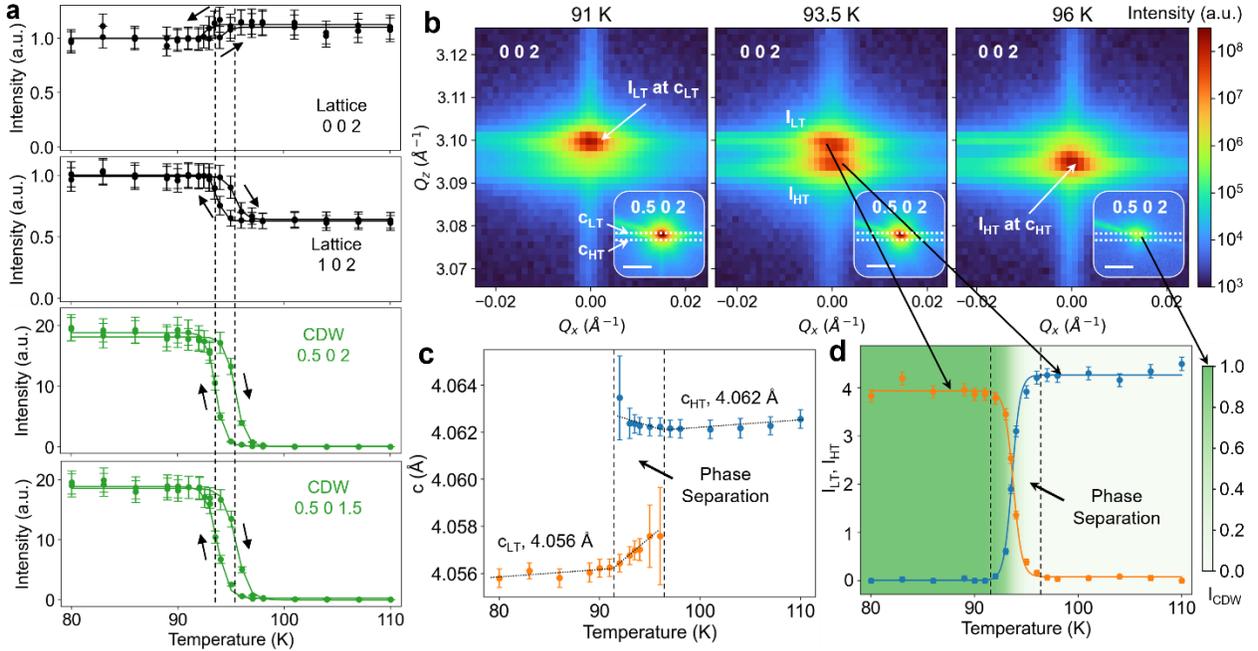

**Figure 2. Structure phase separation revealed by high-resolution synchrotron X-ray diffraction.** (**a**) Integrated intensities of the fundamental reflection 0 0 2 and 1 0 2, and the CO reflections 0.5 0 2 and 0.5 0 1.5, as a function of temperature during heating and cooling. Solid lines show the step-function fit. Black arrows point out the directions of temperature evolution. (**b**) Reciprocal space map (RSM) of the 0 0 2 reflection across $T_{CDW}$. The low-temperature (LT) and high-temperature (HT) phases exhibit distinct $Q_z$ positions. The inset shows the RSM of the CO reflection 0.5 0 2. The dotted lines denote the $Q_z$ values of $2\times 2\pi/c_{LT}$ and $2\times 2\pi/c_{HT}$. The LT and HT peaks are fitted separately for respective intensity and positions. (**c**) Lattice constants of the LT and HT phases as a function of temperature. An abrupt ~0.15% change in the *c*-axis lattice constant is observed. (**d**) 0 0 2 intensity evolution of the LT and HT phases as a function of temperatures, with the CO reflection 0.5 0 2 intensity shown as the gradient background (normalized to its intensity at 80 K). The population inversion between the LT and HT phases is coupled with the intensity change and coincides with the suppression of the CO.

Near $T_{CDW}$, the $c_{LT}$ and $c_{HT}$ lattice parameters did not continuously evolve into one another as has been observed in previous studies on as-grown FeGe [20]. Instead, they exhibited an abrupt split and coexisted over a temperature range of ~92-96K, near $T_{CDW}$. This behavior was a hallmark of first-order structural transitions, characterized by discontinuities in the lattice constants. The peak width of the 0 0 2 along $Q_z$ remained sharp throughout the transition (averaging 240 nm correlation length, after subtracting the instrumental broadening dominated by 0.002° monochromator broadening [28]), indicating that long-range order was retained in both the LT and HT phases. The comparable correlation lengths of the two phases might reflect the length scales of stoichiometric regions established during 320 °C annealing of FeGe [16]. In the meantime, the correlation length of the CDW reflections, averaging 210 nm along $Q_z$, was similar to that of the lattice reflections, suggesting that the long-range CDW forms coherently within the FeGe structural domains.

To identify the structural domain species (LT versus HT phase) where the CDW emerges, the $Q_z$ peak positions of the CDW reflection were compared with the $Q_z$ values corresponding to the lattice constants $c_{LT}$ and $c_{HT}$. The insets in **Fig. 2b** and **Fig. S4** indicated that the CDW peak positions were commensurate with $2\pi/c_{LT}$ rather than $2\pi/c_{HT}$, corroborating that the CDW was hosted within the LT structural phase in FeGe. Along similar lines, **Fig. 2d** and **Fig S3b** showed the fitted areas-under-the-curve intensities $I_{LT}$ and $I_{HT}$ of 0 0 2 and 1 0 2 as a function of temperature, alongside the normalized intensity evolution of the 0.5 0 2 CDW reflection (green shading in the background). Within the temperature range of phase separation and coexistence, the emergence and evolution of the $I_{LT}$ near $T_{CDW}$ precisely coincided with the emergence of CDW intensity. This coincidence between CDW intensity and the LT phase fraction highlighted a strong CDW-lattice coupling associated with the reduced out-of-plane lattice parameter $c_{LT}$.

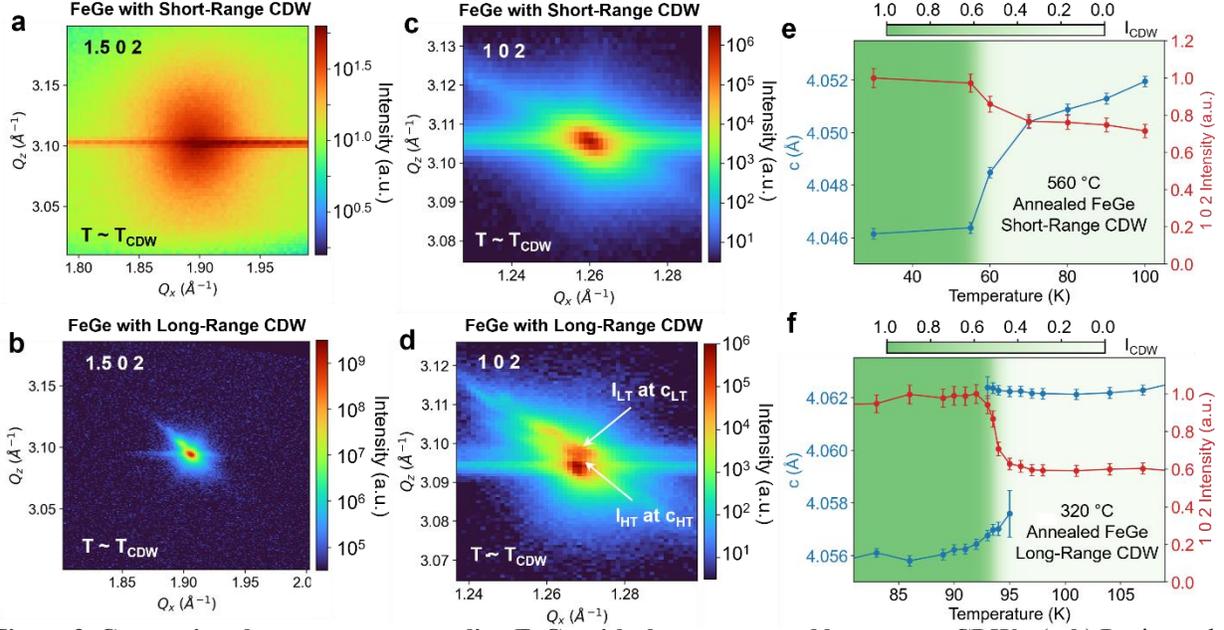

**Figure 3. Comparison between post-annealing FeGe with short-range and long-range CDWs.** (**a-b**) Reciprocal-space map (RSM) comparisons of the CDW reflections between 1.5 0 2 of the (**a**) 560 °C and (**b**) 320 °C annealed (1 0 0) out-of-plane FeGe crystals, shown with identical $Q_x/Q_z$ scale bars. CDW was suppressed in the 560 °C annealed sample until a weak and diffusive CDW became visible below ~55 K. (**c-d**) RSM comparisons of the lattice reflections between (**c**) 560 °C and (**d**) 320 °C annealed FeGe, shown with identical $Q_x/Q_z$ scale bars. Neither doublet peaks nor a broadening of the 1 0 2 were observed in the 560 °C annealed sample in the CDW phase. (**e-f**) Representative intensity and *c*-axis lattice constant as a function of temperature of the (**e**) 560 °C and (**f**) 320 °C annealed FeGe. The normalized CDW reflection intensity is shown as the gradient background.

To further examine the relationship between the lattice transformation, Ge displacements, phase separation, and CDW formation, we performed high-resolution XRD measurements on 560 °C annealed FeGe as a function of temperature (**Figs. 3a, c, e**). We compared it with 320 °C annealed FeGe (**Figs. 3b, d, f**). In the 560 °C annealed sample, a diffuse and broad CDW reflection at 1.5 0 2 was observed, with a coherence length of ~9 nm along $Q_z$ and an intensity approximately five orders of magnitude weaker. Upon heating, the 1.5 0 2 reflection vanished above $T_{CDW}$ ~ 55 K, a temperature much lower than what was observed for the as-grown or 320 °C annealed FeGe ($T_{CDW}$ ~ 100 K). Considering that $T_{SDW}$ is also suppressed to ~25 K in similarly annealed samples [16,22], the strongly reduced CDW order might be linked to an overall lower transition temperatures in 560 °C annealed FeGe.

Comparisons between the long- and short-ranged CDW samples were also made by investigating the 1 0 2 lattice reflections. In the 520 °C annealed FeGe, the 1 0 2 intensity reduced by 28% and its position along $Q_z$ shifted by 0.14%. This response coincided with the emergence of the broadened CDW peak at 1.5 0 2. This response was less pronounced than the larger intensity change and peak shift observed in the 1 0 2 reflection of 320 °C annealed FeGe. The relatively subdued XRD evolution of the 560 °C annealed FeGe upon transition was consistent with a structural picture in which fewer of the Ge1d sites underwent dimerization (estimated to be 0.4 occupancy of dimerized Ge1d sites, as in **Fig. S1f**). This model reproduced the observed CDW reflection intensities more accurately, consistent with previous reports [12,16].

Critically, in the 560 °C annealed FeGe, neither peak splitting nor peak width change of the 1 0 2 lattice was observed throughout the temperature range. Instead, the 1 0 2 reflection shifted continuously (**Fig. 3e**), as opposed to the abrupt lattice constant change and peak splitting (**Fig. 3f**) observed in the 320 °C annealed FeGe. This behavior is analogous to the continuously changing lattice constants previously observed across the transition temperature in as-grown FeGe samples. In these studies, the suggested mechanism was an order-disorder transition mediated by topological defects [20]. Our study revealed that similar outcomes can manifest in systems annealed under specific conditions (560 °C in this case), presenting an avenue for tuning between long-range and short-range CDW ordering.

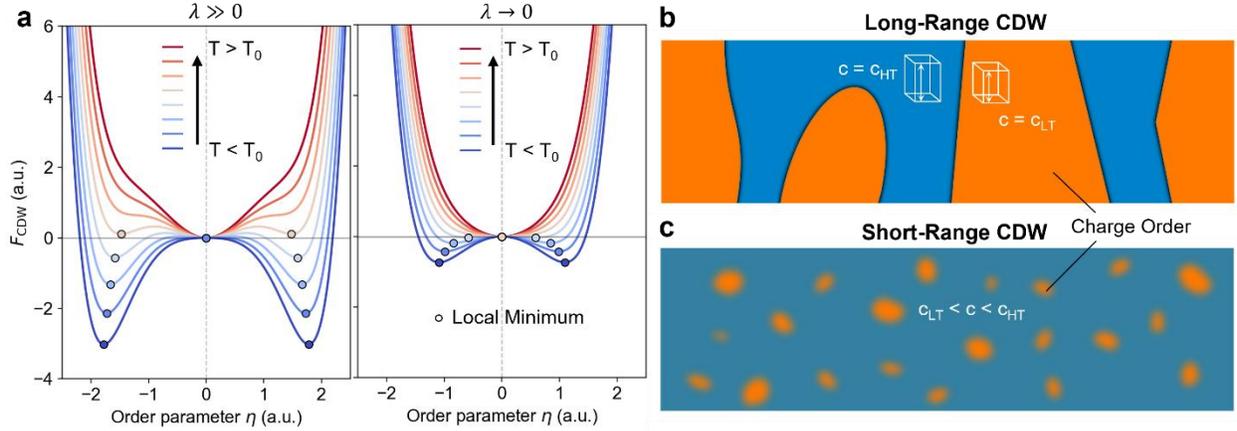

**Figure 4. Qualitative representations of the CDW transition mechanism.** (**a**) The CDW-lattice coupling, $\lambda$, controls the transition mechanism, from a $\lambda \gg 0$ first-order transition in the 320 °C annealed FeGe to a $\lambda \to 0$ continuous transition in the 560 °C annealed sample. (**b**) In the 320 °C annealed FeGe, a first-order phase separation enhances CDW by forming long-range nuclei with a shorter lattice constant $c_{LT}$; (**c**) In the 560 °C annealed FeGe, a continuous transition leads to randomly emerging CO, whose growth is suppressed by distributed Ge vacancies.

To understand the observation of different transition mechanisms in the 320 °C and 560 °C annealed FeGe, we introduce a coupling term between CDW and the out-of-plane lattice constant in a Landau model of the CDW order (details in Supplementary Information, Section 4):

$$F_{CDW} = \frac{A}{2}(T - T_0)\eta^2 + \frac{B}{4}\eta^4 + \frac{C}{6}\eta^6 + \frac{E}{2}\varepsilon^2 + \lambda\varepsilon\eta^2 \quad (1)$$

where $F_{CDW}$ is the change of the free energy due to the emergence of CDW, $A$, $B$, and $C$ stands for the quadratic, quartic, and sextic coefficients of the Landau equation, $\eta$ is the CDW order parameter, $\varepsilon$ is the out-of-plane strain, $E$ is the elastic modulus, and $\lambda$ parameterizes the CDW-lattice coupling. Minimizing with respect to $\varepsilon$ yields equilibrium strain,

$$\varepsilon = -\frac{\lambda}{E}\eta^2 \quad (2)$$

By inserting **Equation 2** into **Equation 1**, the CDW free energy at equilibrium strain is described by

$$F_{CDW} = \frac{A}{2}(T - T_0)\eta^2 + \frac{1}{4}\left(B - \frac{2\lambda^2}{E}\right)\eta^4 + \frac{C}{6}\eta^6 \quad (3)$$

where the effective quartic coefficient can be written as $B_{eff} = B - \frac{2\lambda^2}{E}$. In FeGe, compressive strain ($\varepsilon < 0$) accompanied the emergence of the CDW order. This revealed a strong CDW-lattice coupling in the case of 320 °C annealed FeGe ($\lambda \gg 0$), which was dramatically weaker in the 560 °C annealed FeGe. This strong coupling resulted in a negative quartic coefficient of **Equation 3** ($B_{eff} \ll 0$) in the 320 °C annealed FeGe, resulting in a strong first-order transition character (**Fig. 4a**). Performing this analysis for the case of 560 °C annealed FeGe resulted in a weak CDW-strain coupling, which yielded a strain-insensitive quartic coefficient, $B_{eff} = B - \frac{2\lambda^2}{E} \approx B$. In this case, $B_{eff}$ could result in a second-order or weak first-order transitions.

The positive CDW-lattice coupling also leads to an elevated transition temperature, $T_{CDW}$. For the spontaneous CDW transition from $\eta = 0$ to $\eta \neq 0$, $F_{CDW}(\eta \neq 0)$ has to be smaller than $F_{CDW}(\eta = 0)$. Furthermore, because the local minimum at $\eta \neq 0$ satisfies $\partial F_{CDW}/\partial \eta = 0$, the condition $F_{CDW}(\eta \neq 0) = F_{CDW}(\eta = 0)$ determines the transition temperature (see Supplementary Information, Section 6 for details)

$$T_{CDW} = T_0 + \frac{3\left(B - \frac{2\lambda^2}{E}\right)^2}{16AC} \quad (3)$$

As $B - \frac{2\lambda^2}{E}$ became heavily negative in the 320 °C annealed FeGe under the effect of $\lambda \gg 0$, $T_{CDW}$ became higher than $\lambda \to 0$. This is consistent with our experimental findings that 560 °C annealed FeGe exhibited a significantly lower transition temperature.

The macroscale observations of these transitions discussed above could also be interpreted in terms of different microscopic pictures of the CDW emergence in FeGe (**Figure 4b-c**). In 320 °C annealed FeGe, experimental evidence in other studies [16] suggested that Ge vacancies clustered into >100 nm stripes and created stoichiometric regions [16], allowing the CDW transition to propagate into wider regions without being interrupted by Ge vacancies. This process occurred via a first-order structural transition that nucleated emergent LT or HT phases at $T_{CDW}$. The strain generated by the lattice-constant change is relieved at the domain boundaries, qualitatively represented by the coexisting c = $c_{LT}$ and c = $c_{HT}$ phases in **Figure 4b**.

In contrast, 560 °C annealed FeGe exhibited fundamentally different CDW behavior due to distributed Ge vacancies that suppressed long-range order formation. Near $T_{CDW}$, the transition character we observed was akin to a second-order transition. **Figure 4c** qualitatively illustrates a conceptional inhomogeneous second-order phase transition mechanism. One possible origin of this continuous transition behavior was reported to be topological disorder unbinding between the frustrated CDW [20], which drove continuous lattice evolution and allowed local fluctuations of the order parameter to form short-range CDW correlations [29,30] (**Figure 4b**).

In conclusion, we present direct evidence of structural phase coexistence during the CDW transition in FeGe, establishing a robust first-order mechanism for long-range charge order formation. This degree of CDW-lattice coupling distinguishes FeGe from continuous or weakly first-order CDW systems such as rare-earth tritellurides [31] and $CsV_3Sb_5$ [32]. More strikingly, the resolved splitting of structural Bragg peaks reveals selective CDW coupling to the smaller *c*-axis lattice constant in FeGe. Such structural phase separation is rare even in canonical first-order systems like 1T-$TaS_2$ [33], where first-order character is typically inferred from hysteresis and calorimetry rather than structural discontinuities [34,35]. This unusually strong CDW-lattice coupling positions lattice strain as a key driver of charge order in antiferromagnetic kagome metals.

## ACKNOWLEDGMENTS
This work was supported by the U.S. Department of Energy (DOE), Office of Science (SC), Basic Energy Sciences, Materials Sciences and Engineering Division. The high-resolution XRD was performed on APS beam time awards (DOI: https://doi.org/10.46936/APS-190149/60014218, https://doi.org/10.46936/APS-187978/60012995) from the Advanced Photon Source, a U.S. Department of Energy (DOE) Office of Science user facility operated for the DOE Office of Science by Argonne National Laboratory under Contract No. DE-AC02-06CH11357. The single crystal synthesis and characterization work at Rice is supported by US NSF DMR-2401084 and the Robert A. Welch Foundation under Grant No. C-1839, respectively (P.D.). Y.L. and L.W. acknowledge support from the Department of Energy Early Career Research Program Award under No. DE-SC0026208. Q.D. was mainly supported by the Vagelos Institute of Energy Science and Technology Graduate Fellowship, Dissertation Completion Fellowship, and also partly supported by the National Science Foundation under grant no. DMR-2213891 and the Air Force Office of Scientific Research under award no. FA9550-22-1-0410.

[1] L. Balents, Spin liquids in frustrated magnets, Nature **464**, 199 (2010).
[2] M. Kang et al., Dirac fermions and flat bands in the ideal kagome metal FeSn, Nat. Mater. **19**, 163 (2020).
[3] W. Ma et al., Rare Earth Engineering in $RMn_6Sn_6$ (R = Gd−Tm, Lu) Topological Kagome Magnets, Phys. Rev. Lett. **126**, 246602 (2021).
[4] L. Ye et al., Massive Dirac fermions in a ferromagnetic kagome metal, Nature **555**, 638 (2018).
[5] Y.-X. Jiang et al., Unconventional chiral charge order in kagome superconductor $KV_3Sb_5$, Nat. Mater. **20**, 1353 (2021).
[6] H. Li et al., Observation of Unconventional Charge Density Wave without Acoustic Phonon Anomaly in Kagome Superconductors $AV_3Sb_5$ (A = Rb, Cs), Phys. Rev. X **11**, 031050 (2021).
[7] B. R. Ortiz et al., New kagome prototype materials: discovery of $KV_3Sb_5$, $RbV_3Sb_5$, and $CsV_3Sb_5$, Phys. Rev. Materials **3**, 094407 (2019).
[8] A. Korshunov et al., Softening of a flat phonon mode in the kagome $ScV_6Sn_6$, Nat Commun **14**, 6646 (2023).
[9] X. Teng et al., Discovery of charge density wave in a kagome lattice antiferromagnet, Nature **609**, 490 (2022).
[10] X. Teng et al., Magnetism and charge density wave order in kagome FeGe, Nat. Phys. **19**, 814 (2023).
[11] H. Miao et al., Signature of spin-phonon coupling driven charge density wave in a kagome magnet, Nat Commun **14**, 6183 (2023).
[12] Z. Chen et al., Discovery of a long-ranged charge order with 1/4 Ge1-dimerization in an antiferromagnetic Kagome metal, Nat Commun **15**, 6262 (2024).

[13] L. Wu, Y. Hu, D. Fan, D. Wang, and X. Wan, Electron-Correlation-Induced Charge Density Wave in FeGe, Chinese Phys. Lett. **40**, 117103 (2023).
[14] B. Zhang, J. Ji, C. Xu, and H. Xiang, Electronic and magnetic origins of unconventional charge density wave in kagome FeGe, Phys. Rev. B **110**, 125139 (2024).
[15] S. Shao et al., Intertwining of Magnetism and Charge Ordering in Kagome FeGe, ACS Nano **17**, 10164 (2023).
[16] M. L. Klemm et al., Vacancy-induced suppression of charge density wave order and its impact on magnetic order in kagome antiferromagnet FeGe, Nat Commun **16**, 3313 (2025).
[17] M. L. Klemm et al., Interacting spin and charge density waves in the kagome metal FeGe, Phys. Rev. B **112**, 174422 (2025).
[18] S. Wu et al., Symmetry Breaking and Ascending in the Magnetic Kagome Metal FeGe, Phys. Rev. X **14**, 011043 (2024).
[19] C. Shi et al., Charge density wave with suppressed long-range structural modulation in canted antiferromagnetic kagome FeGe, Commun Phys **8**, 405 (2025).
[20] D. Subires et al., Frustrated charge density wave and quasi-long-range bond-orientational order in the magnetic kagome FeGe, Nat Commun **16**, 4091 (2025).
[21] X. Wen et al., Unconventional charge density wave in a kagome lattice antiferromagnet FeGe, Phys. Rev. Research **6**, 033222 (2024).
[22] J. S. Oh et al., Disentangling the intertwined orders in a magnetic kagome metal, Sci. Adv. **11**, eadt2195 (2025).
[23] S. Yi et al., Charge Dynamics of an Unconventional Three-Dimensional Charge Density Wave in Kagome FeGe, Phys. Rev. Lett. **134**, 086902 (2025).
[24] J. Hammonds, RSMap3D: Reciprocal Space Mapping Software, (2016).
[25] H. Tan and B. Yan, Disordered charge density waves in the kagome metal FeGe, Phys. Rev. B **111**, 045160 (2025).
[26] X. Wu, X. Mi, L. Zhang, C.-W. Wang, N. Maraytta, X. Zhou, M. He, M. Merz, Y. Chai, and A. Wang, Annealing-Tunable Charge Density Wave in the Magnetic Kagome Material FeGe, Phys. Rev. Lett. **132**, 256501 (2024).
[27] C. Shi et al., Annealing-induced long-range charge density wave order in magnetic kagome FeGe: Fluctuations and disordered structure, Sci. China Phys. Mech. Astron. **67**, 117012 (2024).
[28] A. L. Patterson, The Scherrer Formula for X-Ray Particle Size Determination, Phys. Rev. **56**, 978 (1939).
[29] R. Paul, S. Puri, and H. Rieger, Domain growth in Ising systems with quenched disorder, Phys. Rev. E **71**, 061109 (2005).
[30] Y. Feng, J. Wang, R. Jaramillo, J. Van Wezel, S. Haravifard, G. Srajer, Y. Liu, Z.-A. Xu, P. B. Littlewood, and T. F. Rosenbaum, Order parameter fluctuations at a buried quantum critical point, Proc. Natl. Acad. Sci. U.S.A. **109**, 7224 (2012).
[31] N. Ru, C. L. Condron, G. Y. Margulis, K. Y. Shin, J. Laverock, S. B. Dugdale, M. F. Toney, and I. R. Fisher, Effect of chemical pressure on the charge density wave transition in rare-earth tritellurides $RTe_3$, Phys. Rev. B **77**, 035114 (2008).
[32] B. R. Ortiz et al., $CsV_3Sb_5$: A $Z_2$ Topological Kagome Metal with a Superconducting Ground State, Phys. Rev. Lett. **125**, 247002 (2020).
[33] W. Wang, D. Dietzel, and A. Schirmeisen, Lattice Discontinuities of $1T-TaS_2$ across First Order Charge Density Wave Phase Transitions, Sci Rep **9**, 7066 (2019).
[34] Y. D. Wang, W. L. Yao, Z. M. Xin, T. T. Han, Z. G. Wang, L. Chen, C. Cai, Y. Li, and Y. Zhang, Band insulator to Mott insulator transition in $1T-TaS_2$, Nat Commun **11**, 4215 (2020).
[35] A. De La Torre et al., Dynamic phase transition in $1T-TaS_2$ via a thermal quench, Nat. Phys. **21**, 1267 (2025).